# Reflections on the analysis of interfaces and grain boundaries by atom probe tomography

Benjamin M. Jenkins[1], Frédéric Danoix[3], Mohamed Gouné[4], Paul. A. J. Bagot[1,*], Zirong Peng[2], Michael P. Moody[1,*], Baptiste Gault[2,5 *]

[1] Department of Materials, University of Oxford, Parks Road, OX1 3PH, Oxford, UK

[2] Normandie Univ, UNIROUEN, INSA Rouen, CNRS, Groupe de Physique des Matériaux, 76000 Rouen, France

[3] Institut de la Matière Condensée de Bordeaux (ICMCB), CNRS, Université de Bordeaux, France

[4] Max-Planck-Institut für Eisenforschung, Max-Planck-Straße 1, Düsseldorf, Germany

[5] Department of Materials, Imperial College London, Royal School of Mine, Exhibition Road, London, SW7 2AZ - UK

* Corresponding author: **michael.moody@materials.ox.ac.uk, b.gault@mpie.de**

## Abstract

Interfaces play critical roles in materials, and are usually both structurally and compositionally complex microstructural features. The precise characterization of their nature in three-dimensions at the atomic-scale is one of the grand challenges for microscopy and microanalysis, as this information is crucial to establish structure-property relationships. Atom probe tomography is well-suited to analyzing the chemistry of interfaces at the nanoscale. However, optimizing such microanalysis of interfaces requires great care in the implementation across all aspects of the technique, from specimen preparation to data analysis and ultimately the interpretation this information. This article provides critical perspectives on key aspects pertaining to spatial resolution limits and the issues with compositional analysis that can limit the quantification of interface measurements. Here, we use the example of grain boundaries in steels, however the results are applicable for the characterization of grain boundaries and transformation interfaces in a very wide range of industrially relevant engineering materials.

# 1 Introduction

The mechanical properties of metallic materials are usually controlled by their microstructure. During processing, varying parameters allow for changing the grain size distribution, as well as the volume, size and morphology of secondary phases. The composition and structure of interphase interfaces as well as grain boundaries also evolves and has a tremendous influence on physical properties. Knowledge of the precise composition and structure of interfaces has progressively been established via careful microscopy and microanalysis whenever possible at near-atomic resolution. Yet there are still aspects of the true, detailed atomic structure and composition of an interface or gain boundary that remain unresolved. Field-ion microscopy and later atom probe tomography (APT) analyses have significantly complemented extensive transmission electron microscopy (TEM) investigations. The strength of the combination of was further demonstrated by the development of direct correlative approaches [1–4], including at high-resolution [5,6].

APT has risen in prominence as a microanalytical technique over the past two decades, in particular due to its unique combination of compositional sensitivity and capacity for three-dimensional analytical imaging at the sub-nanometer scale [7–9]. Hence, APT would appear perfectly suited for the analysis of interfaces. Yet, APT is primarily a mass spectrometry technique [10], albeit with very high spatial resolution [11–14]. The spatial resolution in APT results from a complex interplay between the field evaporation process that dictates the order in which ions get removed from the surface [13–15], and the shape of the specimen up to the level of the atomic arrangements at the specimen's surface. Combined, these factors determine the nature of the projection of the ions from the apex of the specimen onto the position-sensitive ion detector [16,17]. The simple approach implemented in the commonly used reconstruction protocol [18–20], which generates the 3D atom-by-atom image of the original specimen, completely ignores such complexities. In turn, this strongly limits the accuracy and precision of the analysis of interfaces.

Inaccuracies in the reconstruction associated to trajectory aberrations coming from a specific field evaporation behavior of the interface or grain boundary region can often be identified by fluctuations in the atomic density, i.e. the point density in the reconstructed data [21–23]. These fluctuations can sometimes be used to trace the location of features of interest [24], but most often simply lead to an uncontrolled degradation of the spatial performance of APT. These effects have led to strong debates regarding the accuracy of APT for the characterization of interfaces, particularly in comparison to other microscopy techniques, such as high-resolution (scanning) transmission electron microscopy (HR-(S)TEM) and associated microanalytical techniques such as energy-dispersive X-ray spectroscopy (EDS). HR-(S)TEM often reveals near-atomically-sharp interfaces at grain boundaries in metals [25–27] or interphase interfaces. In contrast, measured APT composition profiles are rarely below several nanometers in width. Such concentration profiles provide integral values over a given area of an interface, and there is evidence that the spatial resolution widely impact the measured profiles [28].

It is also common for researchers, on the basis of APT analysis, to report a single value of the composition of the interface, or more recently the trend is to report the relative excess of solutes [29], following the early work by Krakauer and Seidman [30]. However variations in the local composition at an interface may be revealing of actual physical phenomena pertaining to e.g. segregation or phase transformation [31]. Therefore, the tendency to only report a single value leads to those being overlooked when, for example, correlating the nature of interfaces to resulting material properties.

Here, in the analysis several exemplar and simulated materials systems, we aim to provide some perspective on how the processing of the data itself can cause issues beyond the intrinsic limitations of the technique, in particular when it comes to reporting on the width of a segregation, how the excess might not be devoid of issues, and how those utilizing the APT technique can learn from practices in other communities.

## 2   Materials and methods

In the first section, the material investigated was a ternary Fe-0.12 wt%C-2 wt%Mn, prepared in a vacuum induction furnace. The ingot was hot-rolled, and subsequently cold-rolled. Samples were reaustenitized at 1250 °C for 48 h under Ar atmosphere in order to remove any Mn microsegregation and prevent any decarburization, and finally cold-rolled to a 1 mm thickness. The sample of interest here was heated at 10 °C/s to 1100 °C for 1 min, cooled down rapidly to 680 °C, within the dilatometer, and maintained at this temperature for 3 h (10,800 s). A transformation interface was targeted by using scanning electron microscopy and electron backscattered diffraction (EBSD) to prepare specimens for atom probe by focused-ion beam milling. A bar of the material containing the interface of interest was lifted out, mounted on a support and milled into a sharp needle with suitable dimensions for APT analysis [32]. All the details of the preparation can be found in ref. [33]. APT data was acquired on a Cameca LEAP 4000 HR, at a base temperature of 80 K, in high-voltage pulsing mode with a pulse fraction of 20% and at a repetition rate of 200 kHz. Data reconstruction and processing was performed in Cameca IVAS® 3.6.8.

The material investigated in the second section is a forged ASME SA508 Grade 4N bainitic steel in the quenched and tempered condition. The composition of the bainitic steel is shown in Table 1.

*Table 1: Nominal composition (wt. %) of the ASME SA508 Grade 4N bainitic steel.*

| Element | C | Mn | P | Si | Ni | Cr | Mo | V | Cu |
|---|---|---|---|---|---|---|---|---|---|
| Composition (wt. %) | 0.2 | 0.31 | 0.005 | 0.1 | 3.84 | 1.81 | 0.53 | 0.4 | 0.03 |

A specimen, containing a grain boundary, was prepared for APT analysis using focused-ion beam milling on a Zeiss NVision 40 dual-beam SEM/FIB. Standard FIB procedures were followed [34,35]. APT analysis was conducted using a Cameca LEAP 5000 XR, with a base temperature of 50 K, pulse frequency of 200 kHz and a pulse fraction of 25 %. Data reconstruction was performed in Cameca IVAS® 3.6.8.

## 3 Compositional width of an interface

### 3.1 Background

In steels, the allotropic transformation from the high-temperature face-centered-cubic (fcc) phase to the low-temperature body-centered cubic is one of the degrees of freedom that can be used to adjust the alloy's properties. The partitioning of solutes between the body-centered cubic (bcc)-ferrite and fcc-austenite and their interactions with migrating α-γ interfaces during the growth of ferrite has been a topic of intense research for decades, as recently reviewed thoroughly [36,37]. Modelling the growth of ferrite in low alloyed steels has been extensively investigated because of its great importance for the design of new steel grades [38]. Precise measurements of the local composition of solutes at and in the vicinity of the moving interface are sparse [33,39–42]. Here, we explore how the measured width of the profile is dependent on the local fluctuations of the depth resolution of the technique and that, by selecting the appropriate region, the width of the profile can be in the range of 4 to 5 atomic (011) planes (< 1nm).

## 3.2 Experimental results

Figure 1 (a) shows a tomographic reconstruction containing an α-γ interface, which was selected due to its close adherence to the Kurdjumov-Sachs (K-S) orientation relationship (OR) [33]. Application of the filtering technique introduced by Yao in ref. [43] reveals a clear crystallographic pole in the detector hit maps on both side of the interface, as shown in Figure 1 (b) and (c). The likelihood of observing a pole in the desorption pattern formed on the detector is directly proportional to the inter-planar spacing in this direction. Hence, when only a single pole is observed, it likely represents a low index direction. Here, we made use of the information from the correlative EBSD analysis to guide the identification of the pole as being a (011) in the bcc-ferrite. Upon cooling γ has transformed into martensite and is hence body-centered tetragonal (bct). Assuming that the K-S OR also applied to the austenite-martensite transformation, then the $(011)_{martensite}$ originates from the (111) [44]. This would explain the shape of the pole seen in the bottom grain, which can hence be identified as also being (011) in the martensite. Superimposed poles have previously been considered as an indication of a specific orientation relationship [45]. For this particular interface, the relationship between $(011)_{martensite}$ //$(011)_α$, has been previously reported [46]. With only a single pole visible in each grain, the full analysis of the misorientation cannot be performed from the APT data [47,48]. However, assuming that the angular field-of-view is 55°, the change in the pole position would translate into approx. 2° difference in orientation between the two grains.

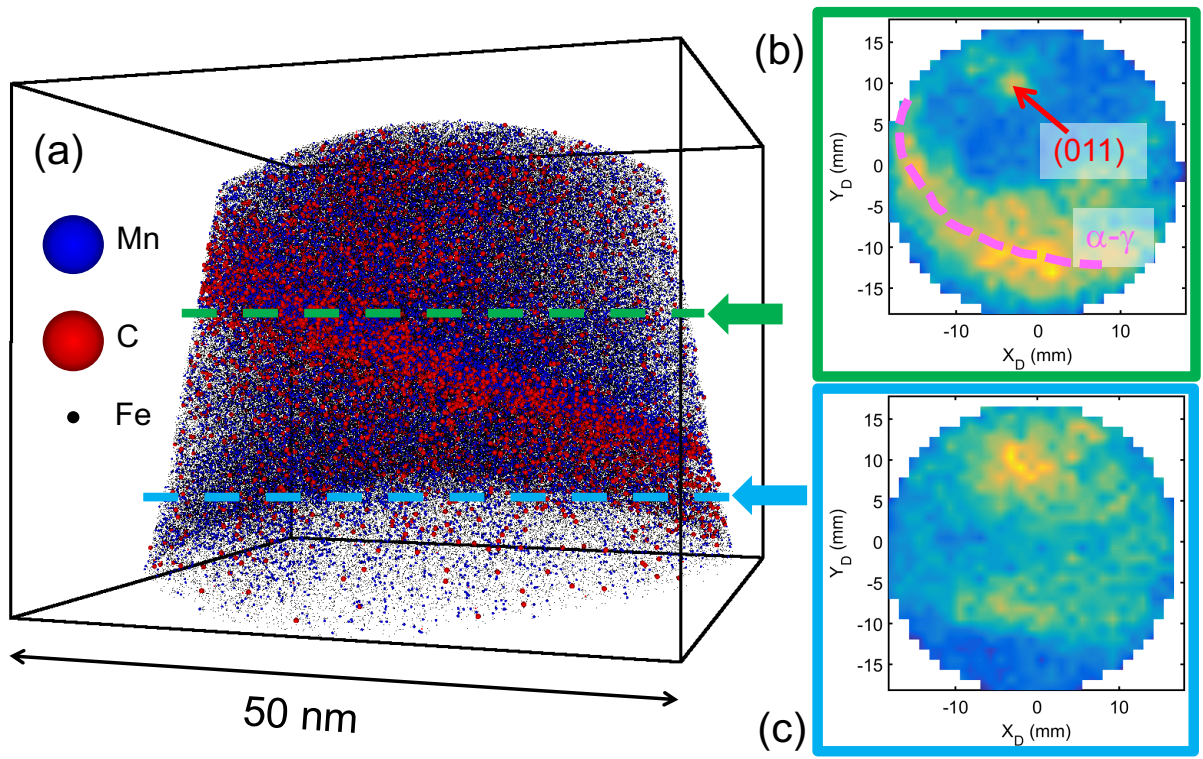

Figure 1: (a) reconstructed APT map showing the distribution of Mn, C and Fe in the dataset containing the interface. For clarity, only 5% of the Fe ions are displayed. (b-c) detector hit maps calculated for a slice of 0.5 million ions at different depths indicated by the arrow of the corresponding color in (a). In (b), a pole is indicated with a red arrow and the position of the α-γ interface is marked by a pink dashed line.

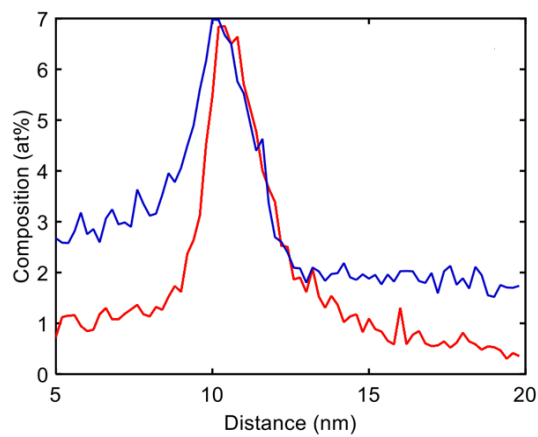

Figure 2: Carbon and manganese composition profiles, red and blue respectively, along a cylinder encompassing the entire interface within the dataset positioned perpendicular to the interface.

Figure 2 shows a carbon composition profile calculated within a cylindrical region of interest that crosses the entire interface, aligned manually as close as possible to normal to the interface. The full-width at half-maximum (FWHM) of the carbon peak across the is approx. 2.3 nm, consistent with previous reports [33,39]. The location of the crystallographic pole in the top and bottom grains indicate where the spatial resolution of this measurement will be maximized. Hence, composition profiles were calculated along a series of 4nm-diameter cylinders positioned at systematically increasing distances from the pole along the interface, as indicated in Figure 3(a). Each profile was then fitted with a Gaussian function to derive the local width and amplitude of the peak. In Figure 3(b) the cumulative number of carbon atoms detected is plotted as a function of the cumulative number of all atoms detected along each of the cylinders. This analysis is known as an integral profile and can provide a measure of the solute excess [30]. The thick purple line is the profile obtained at the pole, and it clearly shows the sharpest transition, which contrasts with the transition observed further away from the pole, e.g. 25nm. Figure 3(c) reports the change in the FWHM of the composition peak obtained from the fitted Gaussian function. At, or near the pole, the FWHM of the peak is in the range of 1nm for both C and Mn. Similar observations of an erroneous widening of thin interfacial layer in the reconstructed APT data as a function of the distance of a pole have previously been reported [49].

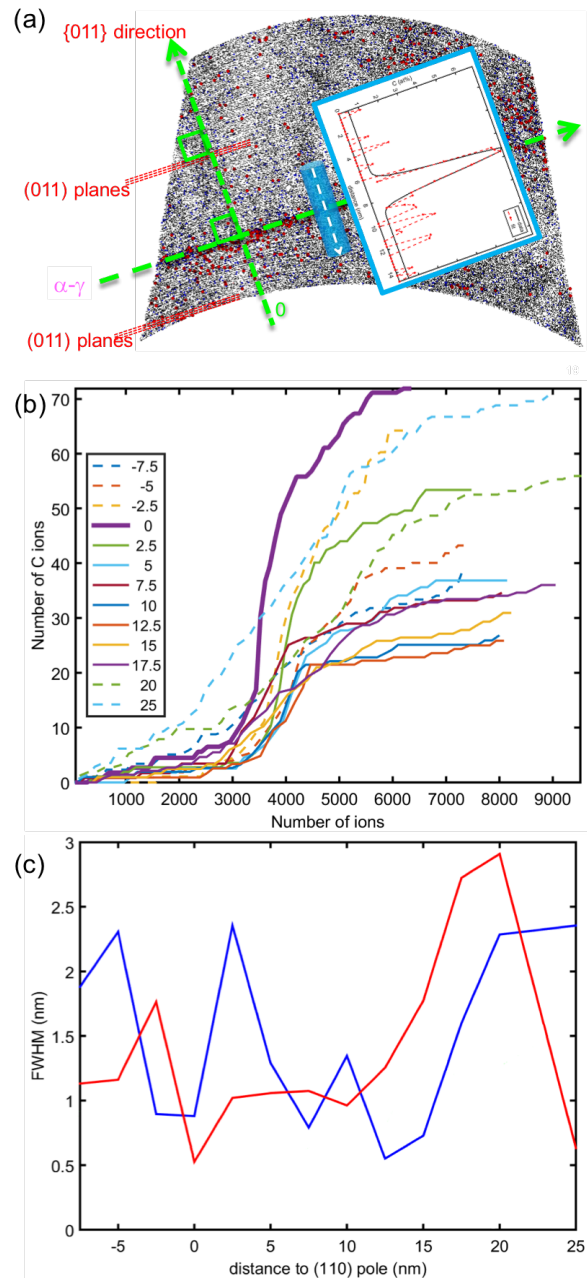

Figure 3: (a) 5nm thick slice through the data showing the interface edge-on and that contains the trace of the two (011) poles and corresponding sets of (011) planes. Two normal axes are defined at the crossing between the interface and the poles. A succession of profiles is calculated within a 4nm-diameter cylinder and each profile is fitted with a Gaussian function, as shown inset. (b) integral profile for each of the corresponding profiles, the color reported in the legend indicates the distance to the pole. (c) full-width-half-maximum of the fitted Gaussian function for the C (red) and Mn (blue).

## 3.3 In-plane solute distribution

These significant changes in the local excess motivated a more detailed investigation of the distribution of solutes at the interface was subsequently undertaken. Figure 4(a) is a plane view of the interface, within a 5nm-thick slice. An iso-composition surface encompassing regions of the APT point cloud where the Mn composition is higher than 6 at% was added. Interestingly, this surface reveals two elongated regions with a high composition of Mn. These appear similar to Mn-decorated dislocations as recently reported [50,51]. These dislocations likely sit at the interface to accommodate the slight misorientation. The distance between the dislocations is approx. 18nm, which according to Frank's equation and for typical Burgers vectors of dislocations on the <110> planes would correspond to less approx. 1 degree misorientation. In Figure 4(b) three 5nm-diameter cylindrical regions of interest are indicated within the atom map, and colored pink, brown and light blue, respectively. The corresponding composition profiles of Mn and C are plotted in Figure 4(c), (d) and (e), respectively. These profiles indicate that there are significant fluctuations of the local composition at the interface, indeed, the peak Mn composition at the dislocations is in the range of 10 at%, while that of carbon is in the range of 8–10at%. These segregations also explain the fluctuations of the excess revealed in Figure 3(b).

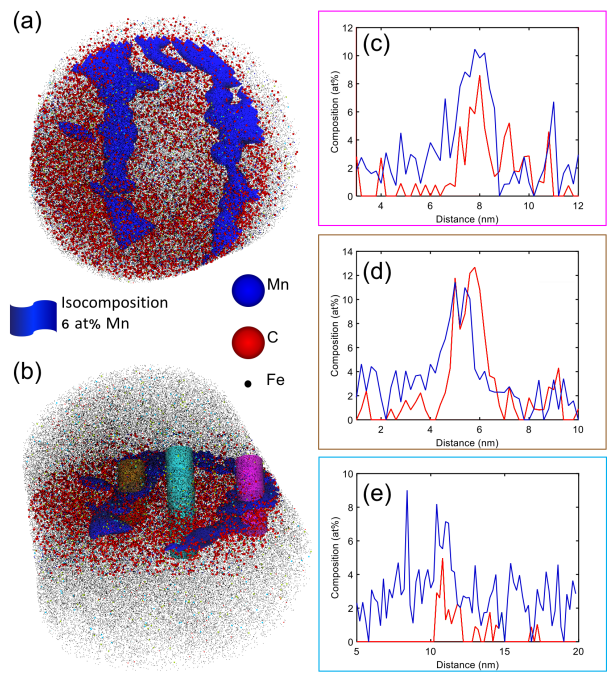

*Figure 4: distribution of the atoms in the plane of the interface, with an added iso-composition surface viewed (a) from the top and (b) tilted to show the three cylindrical regions-of-interest used to calculate the composition profiles in (c–e), each profile bounded by a rectangle of the corresponding colour.*

The Mn segregation at the interface originated from during ferrite growth at 680°C and not at lower temperatures (i.e. during the quench or at room temperature), as the diffusivity of Mn in austenite is already only approx. $10^{-19}$ m$^2$s$^{-1}$ at 680°C [36]. Regarding C, it has been shown to diffuse even at room temperature and C segregation could happen during quench or specimen storage at room temperature (RT) [39]. However, the observed dislocations could carry Mn within the interface and assist the diffusion of C, enhancing the likelihood of carbon diffusing within the interface during the ferritic transformation. Finally, on the basis of thermodynamic arguments, it has previously been shown that the presence of Mn at austenite grain boundaries induces the co-segregation of C [52]. The case of a an α/γ interface is likely more complex, because of the different phases and associated different thermodynamic interactions on either sides of the interface, and the uncertainty associated to the properties of the interface itself. We can however conclude that C segregation to the α/γ interface is likely, provided that a Mn segregation occurs concomitantly during the transformation, strengthening the likelihood of a coupled solute-drag mechanism as suggested in ref. [33].

## 4 Grain Boundary Analysis by APT

### 4.1 Background

It is desirable to be able to quantitatively measure the segregation behavior of elements present at grain boundaries in a reliable and reproducible way. Satisfying both of these criteria is required if multiple measurements of grain boundary segregation are to be used comparatively. This is a necessity if a thorough understanding of how the chemical nature of a grain boundary varies as a result of dissimilar grain boundary physical structure or due to exposure to different environments.

In the case of the ASME SA508 Grade 4N bainitic steel, exposure to elevated temperature for long periods of time was observed to lead to non-hardening embrittlement. Understanding why this non-hardening embrittlement arose is key if models that accurately predict the safe operational lifetime of the component are to be created. It is also of interest to understand grain boundary embrittlement phenomena for the development of new alloys with longer operational lifetimes. Therefore, prior to determining what had caused the embrittlement of the grain boundaries, a reliable, quantitative measurement of the grain boundary chemistry in its as-received state was required. The following section highlights some of the difficulties that arise when attempting to make quantitative measurements from APT data using the most popular and currently implemented analysis method.

Quantitative measures of solute segregation present at interfaces, commonly calculated using the methods proposed by Krakauer and Seidman [30], often reduce the characterization to a single value, i.e. the Gibbsian interfacial excess. However, the previous section highlights some critical challenges when characterizing interfaces using APT, namely: chemical inhomogeneity across this surface; the introduction of subjectivity by requisite user-inputs defining where the interface is sampled and the manner in which the analysis is applied; and inaccuracies originating from reconstruction artefacts. Whilst it is not possible to overcome or account for all of these phenomena, it is important that those interpreting such analyses are aware of them and the potential impact they may have on results. Hence, in the subsequent sections we address some key issues as they relate to the Gibbsian interfacial excess characterization of the chemical nature of a grain boundary using APT.

## 4.2 APT Evaporation Artefacts/Density Changes

A key artefact that has potential to impact the calculation of Gibbsian interfacial excess is the apparent change in atomic density throughout reconstructed atom probe tomography datasets. This can arise due to the presence of crystallographic poles, or due to the different evaporation behavior exhibited by compositionally dissimilar regions.

The measured atomic density (the number of atoms per unit volume) is often higher at interfaces than in the surrounding matrix on either side of the interface (Figure 5). As the grain boundary is likely to have a different composition to the surrounding matrix, the unphysically high measured atomic density at the grain boundary may be the result of the local magnification effect[53]. The amplitude of the local magnification effect at interfaces has been shown to be minimised when the interface is perpendicular to the analysis direction during field evaporation [54]. Furthermore, the variation in atomic density between a grain boundary and the surrounding matrix has previously been shown to arise in grain boundaries which undergo simulated field evaporation [23]. The authors observed that the change in atomic density at the grain boundary occurred even in simulated materials with homogeneous evaporation fields, indicating that the measured atomic density within APT reconstruction is affected by the structural defect of a grain boundary as well as by the varying evaporation fields of the elements present at the boundary [23].

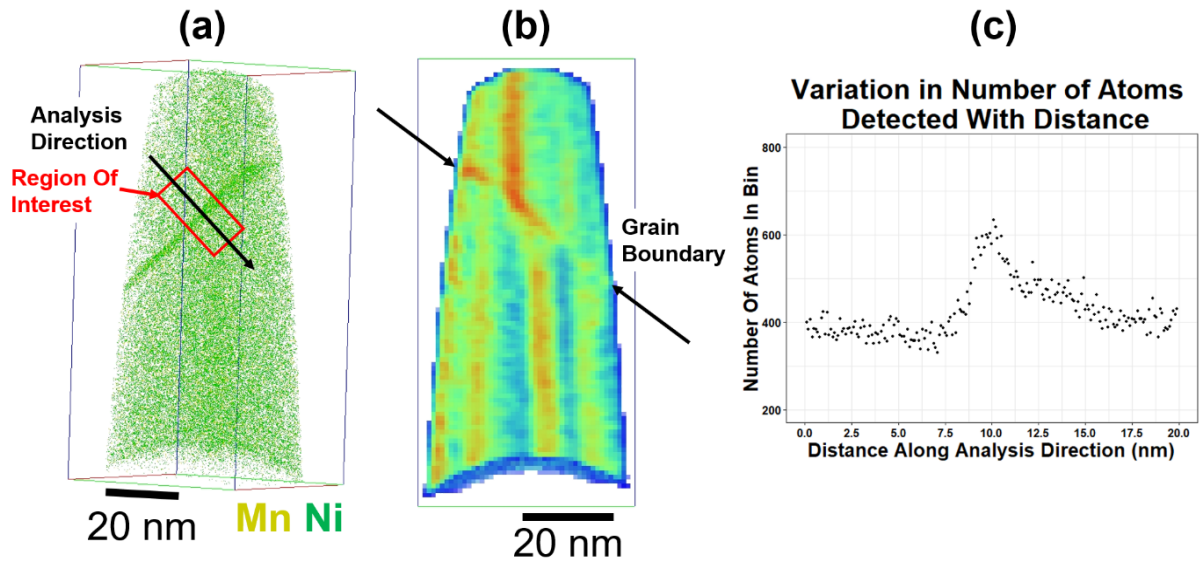

Figure 5: (a) Distribution of Mn and Ni atoms within the specimen. (b) Atomic density (atoms/nm$^3$) measured throughout the APT tip and (c) Variation in number of atoms detected in per 0.1 nm bin along the region of interest in (a).

Both of the above effects lead to more atoms of each species being erroneously reconstructed at the boundary. Therefore, an apparent excess of all atoms would be observed at the boundary even in a homogeneous material. If one were to simply measure the number of atoms of element $i$ within a series of sampling bin (of fixed width) across the interface versus distance, it may appear that there is an excess of $i$ atoms when, in reality, $i$ shows no segregation to the interface. To avoid the above phenomenon, it is important to calculate the Gibbsian interfacial excess by carefully applying the equations outlined in [55]. The aforementioned aberration effects will also lead to the apparent composition of the interface region being different to the true composition. If one is to report composition it is, therefore, important to correct for this [56].

### 4.3 Repeatability of Measurements

A factor that greatly affects the reproducibility of Gibbsian interfacial excess calculations is the large number of parameters that must be selected by the user performing the analysis. These parameters include defining the extents of Grain A and Grain B, the position of the Gibbs dividing surface, the area of the interface that is analyzed, the location the measurement is performed on the interface, and the bin size selected. Varying either of these can have a large effect on the calculated excess values and it is important that users report the parameters that were used, why they were selected, and how sensitive their results are to changes in the parameters.

Another issue that researchers often fail to account for is the consequence of the region of interest not being perpendicular to the interface. If the analysis direction is not perpendicular to the interface, the calculated value of $\Gamma_i$ will underestimate the true value. This underestimation in $\Gamma_i$ arises as, due the contribution of some of the matrix at all distances along the region of interest, the measured peak composition of segregating species will be lower than if the analysis is performed perpendicularly to the interface.

The positions where Grain A ends and the interface begins, and where the interface ends and Grain B begins, respectively, are almost always not clearly defined in experimental data with the interfacial region often taking the form of a tanh function (Figure 6). Whether this shape is the reflective of the true solute distribution, or arises due to aberrations cannot be determined. The user performing the analysis must make a subjective decision as to where to define the positions of these boundaries.

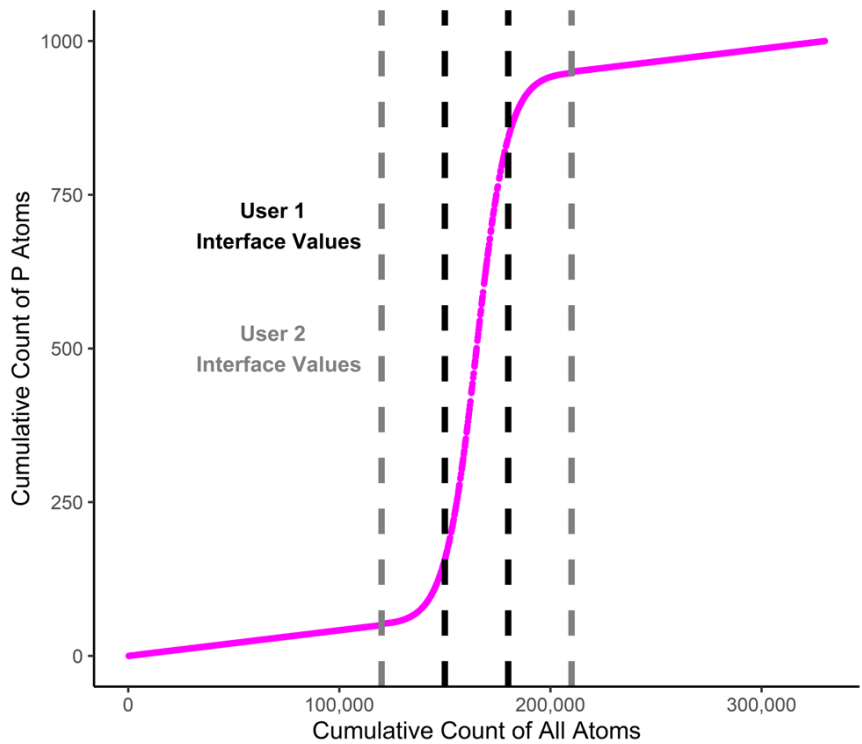

*Figure 6: Simulated cumulative plot of number of P atoms vs cumulative number of all atoms, showing where two users could define the interface region as beginning and ending.*

Selection of the positions defining the extents of the grain boundary can significantly impact the calculation of $\Gamma_i$. Figure 6Figure 6 demonstrates the way in which two independent users may define the position of the grain boundary encountered in Figure 6Figure 6.

The effect this has on the calculated $\Gamma_i$ is not trivial, as demonstrated by the resulting measurements presented in Table 2Table 2. Reducing the sensitivity of the calculated $\Gamma_i$ with respect to user input is therefore critical to make such measurements meaningful and robust. The fitting refinement procedure used in [57], removes the requirement for user input in defining the start and end of the interface region. Other statistical approaches may also be implemented.

| User | Interface Start | Interface End | $\Gamma_P$ (Excess Atoms/nm²) |
|---|---|---|---|
| | | | |

|   | (Cumulative Number Atoms) | (Cumulative Number Atoms) |      |
| - | ------------------------- | ------------------------- | ---- |
| 1 | 150,000                   | 180,000                   | 17.7 |
| 2 | 120,000                   | 210,000                   | 24.1 |

*Table 2: The effect the selected interface start and end values can have on the calculated Gibbsian interfacial excess values (Figure 6 and assuming area = 100 nm² and η = 0.37).*

The location of the Gibbs dividing surface within the interface region will also affect the calculated $\Gamma_i$. However, lower and upper bounds can be determined by placing the surface at the very start or end of the interface region.

### 4.4 Loosely Defined Variables

Another issue which influences the reproducibility of results, is that the definitions provided in the original paper [30] are not strict, particularly in the case of more complex material systems. For example, the authors define $C_i^\alpha$ and $C_i^\beta$ as "the atomic compositions of element $i$ in the homogeneous regions of phases α and β, i.e., the bulk regions of the two phases." However, the segregation of solutes to interfaces can lead to a denuded zone around the interface [58], meaning that phases α and β are not homogeneous. Furthermore, precipitate or cluster formation occurs in many material systems and means that individual grains/phases are often not homogeneous.

If this occurs, then what is precisely meant by the definition of the "homogeneous regions of phases α and β" is no longer rigid. Figure 7Figure 7 demonstrates four different regions in the same material which may be considered "homogeneous" by a user who is calculating Gibssian interfacial excess values. The selection of either of these regions has the potential to greatly affect the calculated $\Gamma_i$ values and, therefore, the reproducibility of results.

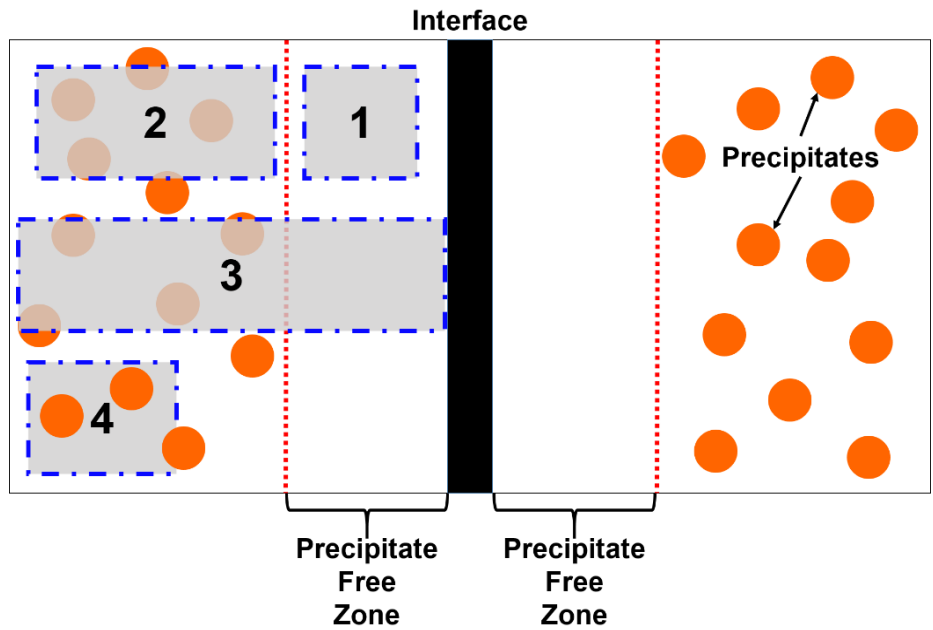

*Figure 7: Schematic diagram demonstrating presence of a precipitate free zone adjacent to an interface, and the four different regions that independent users could decide best reflect the "homogeneous" regions of phase α.*

In Figure 7Figure 7, Region 1 may be considered "homogeneous" as this is the area adjacent to interface and contains no other phases. However, the choice of Region 2 may also be justified since this region is representative of phase α before the precipitate free zone formed. Region 3 could be selected as it samples both the region adjacent to the interface and phase α away from the interface, offering a compromise between Region 1 and Region 2. Region 4 selects the matrix of phase α away from the interface, but does not incorporate the precipitates in the matrix. This choice could be justified, since the precipitates may be a different phase to the α matrix. There is no widely accepted protocol on how to proceed in this circumstance, and a user could reasonably select any of the regions to describe the homogeneous region of phase α. It is, therefore, important that one justifies why and accurately describes how measurements have been made.

## 4.5 Inhomogeneous Interfaces

In the cases of interfaces where solutes are not homogeneously distributed across the interface, the reporting of a single value to describe the segregation leads to a loss of information. Some studies have applied a mesh to the interface to be analyzed and reported a map that shows the variation of $\Gamma_i$ across the boundary [29,57]. This is an improvement, but $\Gamma_i$ will still vary for each region depending on the size of the mesh, another variable which must be defined by the operator. The maps also provide a qualitative, not quantitative description of segregation; this presents an issue in developing mathematical models which simulate GB segregation.

# 5 Discussion

## 5.1 Targeted specimen preparation

It is now routine to combine APT with electron microscopy techniques, in particular electron back-scattered diffraction (EBSD) or electron channeling contrast imaging [59,60] prior to FIB milling in order to select a specific orientation. There are also possibilities to use such techniques during preparation with for example transmission Kikuchi diffraction [61,62]. The preparation of specimens along particular orientations should, when possible, help maximize the spatial resolution, with the optimal configuration being when the interface is strictly perpendicular to the specimen's main axis to limit distortions associated with the tomographic reconstruction. These aspects have been discussed previously but are not commonly taken into account [4].

In the analysis of complex interfaces by TEM-based techniques, the challenge is often to find a suitable orientation to visualize the interface edge-on. This has often led to the use of specific bicrystals or model interfaces, which may not have relevance to microstructures encountered in engineering materials. Analyzing interfaces and grain boundaries with near-atomic resolution by TEM-based techniques, in particular scanning-TEM, requires the two grains to have a common zone axis direction that is close to the normal of the sample surface so as to observe the interface edge-on. The possible broadening of the electron beam travelling through the specimen and the possibility that the interface is not straight, which is likely for transformation interfaces such as the one investigated herein, imposes the use of very thin specimens, in the range of 10–30 nm. The width of this same interface measured by energy-dispersive X-ray spectroscopy in an aberration-corrected STEM is also in the range of several nanometers [33] and so was that measured by electron-energy loss spectroscopy on model interfaces [42]. We demonstrated above once again the importance of maximizing the spatial resolution, which can be done by ensuring that a set of low index atomic planes is close to the center of the field-of-view.

## 5.2 Issues inherent to data processing

The analyses in Section 3 also point to a number of shortcomings of the typical approaches used to extract information from the APT reconstruction. The use of composition profiles as a function to the distance to a selected iso-composition surface (i.e. proximity histogram) has now become widespread [63]. Although the concept of such calculations is interesting, its implementation is not without idiosyncrasies. In particular, this approach requires an isosurface, which is calculated on a grid, which is usually smoothed by a Gaussian blurring function, in a process coined delocalization [64]. This can lead to a strong smoothing of the compositional field and a widening of the actual interface, which is often noticed in the analysis of large populations of precipitates of varying sizes [65]. Alternative approaches have been proposed that may alleviate these concerns [29,31,57], but they are not accessible to most, and they systematically require input parameters. Albeit more labor-intensive, using simpler means of data extraction, e.g. composition profiles, often leads to a better understanding of the underlying assumptions made to obtain information. Here, similarly, scientists using APT must understand the limitations of the technique but also potentially accept not to do what is easy, but limit their analysis to regions in the point cloud that are highly-resolved, which may require finding a suitable orientation and location to analyze the data more deeply.

The results in Figure 3 and Figure 4 point to the importance of performing two-dimensional mapping of the distribution of solutes at interfaces. This has been discussed in several studies recently [29,31,57,66,67], and tools are becoming more easily available. These tools usually allow for compositional mapping but do not yet include means to see if the observed fluctuations are beyond what would be expected in a random distribution of solutes confined to an interfacial region. These tests are commonly applied in the analysis of APT data [68] but have so far not been applied in a two-dimensional case.

Although the information from APT is primarily compositional, often structural information is buried in the data [69]. This has been known since the inception of atom probe tomography, with early reports of atomic planes and segregation to crystalline defects [70]. Through appropriate processing, this information was exploited to push the analysis further. In the investigation of the ternary Fe-0.12 wt%C-2 wt%Mn in Section 3, this was complemented by electron microscopy [50]. Ignoring this information can lead to misinterpretation of the data, whereas it could be crucial to understand microstructural evolution. The values of the composition of Mn and C at the dislocations imaged herein are 20–30% higher than the peak value reported in Figure 2. Solutes are known to pin dislocations, and the presence of such high concentrations of Mn and C at these will affect their mobility. The interface analyzed here is a moving interface, and to accommodate the progressive displacement of the interface, these dislocations likely need to move. The presence of such high compositions needs to be accounted for in models developed to explain the mobility of these transformation interfaces.

Finally, there have been preliminary reports of trying to correct composition for changes of the atomic density [71,72], but these are not widely used and do not correct according to the respective field evaporation behavior of different features. An approach using input from field evaporation simulations was also proposed for precipitates [22], but has not been used for interfaces.

## 5.3 Is the Gibbs Excess Measurement Fit for Purpose?

As well as the issues arising when trying to calculate the Gibbsian interfacial excess from APT data, the validity of using the Gibbsian interfacial excess to correlate changes in properties with the evolution of grain boundary nature in real material systems is debatable.

The quantity $\Gamma_i$ by itself is not a true measure of the nature of the interface. It is a measure of the composition of the interface with respect to the composition of two phases either side of it (i.e. segregation strength).

Consider a simplistic scenario where two batches of material are produced. One batch may have a higher overall impurity ($C_i$) level than the other (Table 3), but the segregation behavior of this impurity element to grain boundaries may be different in each system. If grain boundaries from each of these materials were then analyzed, the composition profiles shown in Figure 8 may be collected.

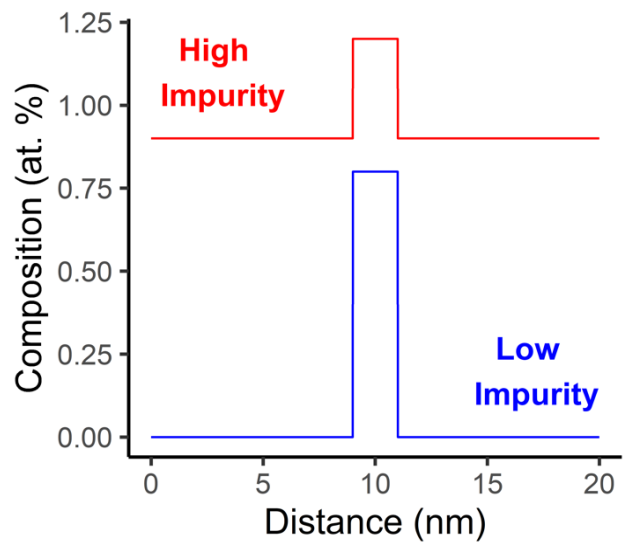

*Figure 8: Cartoon composition profiles across the same type of interface in two respective batches of the same material with different impurity levels.*

It is clear that there is a higher composition of the impurity element, $i$, at the interface in the 'bad batch' material. However, Table 3 shows that the calculated value of $\Gamma_i$ is actually higher for the 'good batch'.

| Batch | $C_i$ (at.%) | $C_i^\alpha$ (at.%) | $C_i^{Boundary}$ (at.%) | $\Gamma_i$ (Excess Atoms/nm²) |
|---|---|---|---|---|
| Good | 0.08 | 0.00 | 0.80 | 8.0 |
| Bad | 0.93 | 0.90 | 1.20 | 3.0 |

*Table 3: Table showing the variation in composition of different regions in two batches of a material, as well as calculated $\Gamma_i$ values (assuming $\eta=1$, $A=1000$ nm², $N=100,000$ atoms, and $\xi=0.5$).*

This raises the question as to what is more important in determining macroscale material properties, the composition of the interface, or the composition of the interface with respect to the matrix. If it is the composition of the interface that is of most importance, then the validity of applying $\Gamma_i$ to relate the character of microstructural interfaces to material properties is questionable. Reporting the Gibbsian interfacial excess of each element, together with the composition of phases α and β would provide a more holistic description of the interface.

A key assumption made by Gibbs in his model was that the interface is a 2D plane [73]. This is likely not strictly true for many interfaces in reality. Guggenheim treated the interface as an interphase with a finite thickness[74]. Since it is known that most grain boundaries do not take the form of idealized 2D features, assuming all segregation is confined to a single plane is likely naive. If this assumption is made, interfacial excess values higher than those permitted by the atomic density of the material are possible. This may be evidence for more than one monolayer of coverage at the interface, however, there is no way to confirm the lattice site location of the excess atoms in the enriched region.

The estimated width of the interface is also extremely important because it determines the transformation kinetics derived from models, e.g. coupled solute drag, reviewed for instance in ref. [36]. It is also related to the binding energy for solute at a moving interface, which is usually derived from such profiles [33,39]. An overestimation of the interface's width leads to an underestimation of the solutes' segregation energy at the interface. Here, by going further into the processing of the data, and targeting regions from within the data where the resolution is optimal, the width of the interface can finally be accurately measured and reported.

The use of the Gibbsian interfacial excess values to calculate thermodynamic quantities relies on the assumption that the system is at thermodynamic equilibrium. However, many materials subject to APT are not at thermodynamic equilibrium at the time of analysis. Therefore, $\Gamma_i$ should not be used to calculate thermodynamic quantities. In the case of non-equilibrium segregation, the segregation will be to a zone "considerably greater width around the appropriate interface than occurred with the equilibrium mechanism…"[75]. The authors state that this zone may vary in thickness from the nanometer to micrometer scale[75]. Therefore, assuming all of the segregation is confined to a single plane does not accurately reflect what is physically present in the system and thereby will lead to an overestimation of the interfacial excess.

## 6 Conclusion

To conclude, we wanted to provide some perspective on the analysis of transformation interfaces and grain boundaries by APT. We showed that how the spatial resolution of the technique affects the width of composition profiles, including within a single dataset. This allowed us to reveal segregation of Mn and C within only less than 0.5–0.6 nm, i.e. two to 3 (110) interplanar spacing. When analyzed appropriately, the data reveals that the transformation interface is only semi-coherent and contains dislocations that lead to a complex segregation behavior with stronger segregation at the dislocations than at the interface. These details had not been revealed before. Other microscopy techniques tend to optimize the specimen preparation strategy to ensure that the desired observation can be performed, and if this is sometimes done in APT, it is not always common practice. We also discussed in detail how the sometimes-blind use of the interfacial excess in lieu of the interfacial composition can lead to details of the analysis being lost. In line with other recent work, we challenged the belief that the excess is not affected by trajectory aberrations, but also provided some discussion point regarding whether the interfacial excess is always an appropriate metric in the case of complex interfaces where phase transformation has occurred. We expect that these points help start a discussion within the community.

## Acknowledgements

BMJ and MPM would like to acknowledge financial support from EPSRC EP/P005640/1 and EP/M022803/1. BMJ and MPM would also like to thank Rolls-Royce Plc. for financial support and for providing the ASME SA508 Grade 4N bainitic steel.